\documentclass[aps,pra,twocolumn,superscriptaddress,nofootinbib,a4paper]{revtex4-2}

\usepackage[pdftex]{graphicx}
\usepackage{amsmath}
\usepackage{verbatim}
\usepackage{xcolor}
\usepackage{bm}
\usepackage{txfonts}
\usepackage{siunitx}
\usepackage[colorlinks=true,allcolors=blue]{hyperref}

\begin{document}
\title{Simulating the electrostatic patch force in experimental geometries}
\author{Matthijs H.\ J.\ de Jong} 
\affiliation{Department of Applied Physics, Aalto University, FI-00076 Aalto, Finland}

\author{Laure Mercier de L{\'e}pinay} 
\affiliation{Department of Applied Physics, Aalto University, FI-00076 Aalto, Finland}

\begin{abstract}
Potential patches are responsible for a force between closely-spaced objects that forms a parasitic contribution to sensitive force measurements. Existing analytical models cannot account for the patch force in the 3D geometries of real experiments. Here, we present a finite-element method model to evaluate the impact of patches in geometries with roughness, edges, and curvature. First, we test our model against the plate-plate and sphere-plate geometries, for which the exact solutions are known. Then, we apply it to more complicated geometries for which analytical solution are challenging, and finally we extend it to handle AFM-measured rough surfaces. Patch textures are generated as a Voronoi diagram representing crystalline grains, or may be imported from potentials measured in Kelvin Probe Force Microscopy experiments. This work provides a reliable estimation of the parasitic contribution from random potential patches in realistic experimental geometries, which may be of relevance to Casimir force measurements or gravitational wave interferometers.
\end{abstract}

\date{\today}
\maketitle


\section{Introduction}
The surface of real-world materials features spatial fluctuations of the effective electrostatic potential known as potential patches~\cite{Camp1991}. These patches can arise from local variations in the work function due to grain orientation~\cite{Sadewasser2002}, adsorbates~\cite{Yurtsever2017}, or defects~\cite{Glatzel2003}. These potential patches exert a force even after the mean potential difference between surfaces is cancelled, and thus they produce a parasitic signal in precision force measurements. Potential patches affect gravitational wave detectors~\cite{Speake1996,Robertson2006,Pollack2008,Vitale2024}, tests of general relativity~\cite{Everitt2011}, inverse-square law experiments~\cite{Antonucci2012,Behunin2014,Tan2020,Dong2023}, exclusions of non-Newtonian forces~\cite{Sushkov2011,Klimchitskaya2012}, Casimir force experiments~\cite{Speake2003}, AFMs~\cite{Dorofeyev1999}, and particles suspended in traps \cite{Turchette2000,Dubessy2009}. 

The patch contribution has been estimated to contribute $<\!1\%$ of the force signal in extremely careful measurements of the Casimir force~\cite{Decca2003,Decca2005,Decca2007,Behunin2012,Zou2013,Garrett2015,Garrett2020}. This magnitude has to be placed in the context of a discrepancy between the measured and predicted Casimir force between real (lossy) metal objects~\cite{Mostepanenko2021}. The parasitic patch force can be statistically estimated and was found to be significant on the level of the discrepancy, but fell short of bridging the difference between the theory and observations. Nonetheless, recent proposals~\cite{Norte2018,Bimonte2019} face tight constraints on the accuracy of estimating the parasitic patch contribution. While potential patches can be measured using Kelvin Probe Force Microscopy (KPFM)~\cite{Garrett2015} and reduced by surface treatments~\cite{Xu2018,Garrett2020}, these methods are often not available in situ~\cite{GarciaSanchez2012}. In this context, new undertakings in proximity force measurements should systematically be paired with a detailed investigation of the patch contribution in the experimental geometry used.

Thus, a large body of work has focused on modeling the effect of potential patches. The seminal paper on estimating the force from potential patches, \cite{Speake2003}, provides an analytic model based on spatial voltage correlations in and between two flat planes~\cite{Behunin2012,Kim2010}. This treatment has been extended to sphere-plate geometries using the proximity force approximation (PFA)~\cite{Fosco2012} and transposed to a bi-spherical coordinate system to describe the same geometry without using the PFA~\cite{Behunin2012a}. However, despite its relevance in Casimir force experiments, this last model has hardly been applied to experimental situations because of its complexity. Additionally, experiments have shown that local imperfections in sphere curvature must be considered~\cite{Kim2008,deMan2009}, and, furthermore, experiments studying the impact of Casimir forces in the context of cavity optomechanics involve more complicated geometries~\cite{Norte2018,deJong2025,Xu2025} are not suitable for analytic treatments. As a result, there is a growing interest for a practical and applicable simulation method to predict the patch force in a given geometry. Recently, a finite-element model was created that generates a random patch realization in a plate-plate geometry~\cite{Ke2023}, but there remains a significant challenge in extending this method to arbitrary 3D geometries, which are more relevant in practice.

In this work, we develop a method to accurately estimate the parasitic contribution of patches in any experimental geometry. With input of patch properties from KPFM measurements on flat pieces of material, we can synthesize patch textures and apply them to the 3D surface of interest. To reach this goal, we create a finite-element model of randomly distributed potential patches that can handle non-planar (e.g., sphere-plate) and more complex geometries, and make it publicly available (see \hyperlink{Paravailability}{Data availability}). We then study the scaling of the patch force with distance in typical geometries and make general recommendations to reduce the parasitic contribution of patches.

\section{Results}
\subsection{Patch simulation model}
\begin{figure*}
\includegraphics[width = \textwidth]{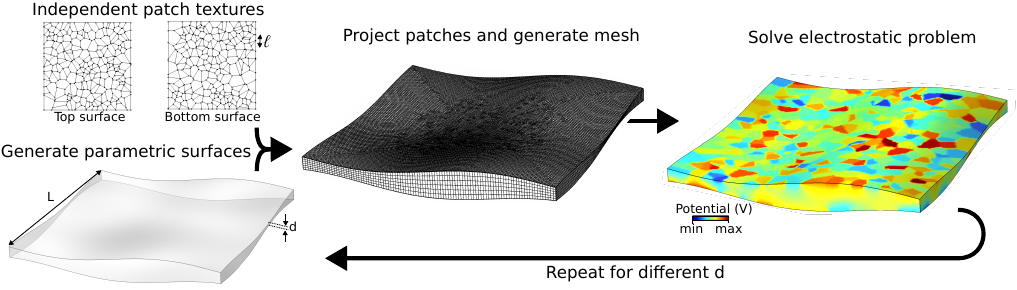}
\caption{Schematic of the patch simulation method. Two independent Voronoi patterns representing patch textures are generated. They are projected onto the top and bottom surfaces of a square volume bounded by two parametric surfaces representing the experimental geometry. Each patch is assigned a random, normally distributed potential value. The finite element mesh of the surface interpolates the potentials from both of the patch textures, and a swept mesh is used for the volume. The electrostatic problem is solved and the total energy is extracted for a given distance $d$. The model then iterates over all desired distances to extract the distance-dependence.}
\label{FigSimulationschematic}
\end{figure*}
We show the working of our model schematically in Fig.~\ref{FigSimulationschematic}. The model geometry is defined by two parametric surfaces (top and bottom) in a square of side length $L$. We generate two independent patch textures as a Voronoi pattern with a controllable number of random patch centers, to mimic crystalline grains~\cite{Ke2023}. We control the number of patches, $n$, which yields the characteristic patch size $\ell = L/\!\sqrt{n}$. Each of the patches is assigned a random potential value, distributed normally with some standard deviation $\sigma_V$. The patch textures are turned into a $256\times 256$ grid of potential points, which is projected on parametric surfaces which form the simulated experimental geometry. We apply a zero-charge boundary condition on the side walls (no electric flux through the boundary), which means we envisage the top and bottom surfaces as parts of larger objects that extend uniformly outwards. The smallest distance between the surfaces is the minimum separation $d$. We solve the electrostatic finite element problem between the two projected patch textures for different separations between the surfaces. 

In the following, unless otherwise specified, we model $1156$ patches with characteristic size $\ell = 10$~\si{\nano\meter} and potentials following a normal distribution of standard deviation $\sigma_V = 100$~\si{\milli\volt}. The lateral extent of the simulation domain is $L = 340$~\si{\nano\meter}, the minimal separation distance is varied between $d = 0.1$~\si{\nano\meter} and $d = 100$~\si{\nano\meter}. The dimensions are chosen such that the curvature of a sphere with radius \SI{250}{\nano\meter} is reasonably well-captured in our model while avoiding reaching the side(s) of the sphere where its surface is orthogonal to the plate; this is a limit of the current implementation of the FEM model. We use a deterministic mesh generation algorithm to mesh the bottom surface, and then use sweeping to generate hexahedral volume elements from this bottom surface mesh. Each vertex on the top surface is thus connected to a vertex on the bottom surface through some arc, and we generate the vertices of the hexahedral elements at explicit points along the (normalized) arc length. For the smallest value of $d$, there are 6 equally spaced elements along the arc. For each subsequent larger value of $d$, we insert 2 new elements in the center, such that the previous elements on each side keep the exact same distance to their respective closest surface as for the previous value of $d$. Hence the elements close to the surface are exactly the same for all values of $d$, which significantly reduces mesh-based variations of the simulation results.


The schematic shown in Fig.~\ref{FigSimulationschematic} projects the patches from a flat plane onto a curved surface, which is acceptable when the surfaces are sufficiently flat and smooth. To respect the average patch size on sharper and more curved geometries, one can adjust the positions of the patch centers to correct for the projection onto the parametric surface. We discuss the limits and possible extensions of this work in later sections.

\subsection{Patch force in plate-plate and sphere-plate systems}
\begin{figure}[ht!]
\includegraphics[width = 0.49\textwidth]{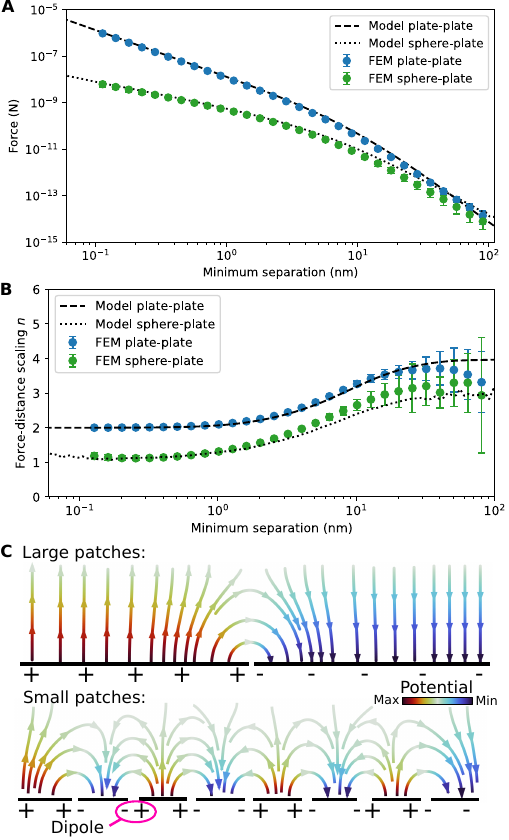}
\caption{\textbf{A}: The pressure exerted by potential patches between two plates (blue), and a sphere-plate (green) as evaluated by our finite element method (FEM) and the numerical evaluation of the analytical expressions. \textbf{B}: Force-distance scaling exponent $F \propto d^n$ for the patch force between two parallel plates (blue) and a sphere and plate (green), evaluated in our FEM model and by the numerical evaluation of the analytical expressions. The errorbars denote the standard deviation of the scaling over $15$ random patch textures. \textbf{C}: Schematic of electric field lines between potential patches in a parallel-plate system (top plate not shown). In the limit of large patches, $\ell \gg d$, the interaction between patches on the same plate is small compared to the interaction of the patches with the other plate. By contrast, in the limit of small patches, $\ell \ll d$, the interaction between patches on the same plate is much stronger than the interaction with the other plate.}
\label{FigSphereplatecomparison}
\end{figure}

The force arising from potential patches can be computed analytically for some simple geometries like two parallel plates~\cite{Speake1996} and a sphere close to a plate~\cite{Behunin2012a}. To validate our FEM model, we generate $15$ randomized patch textures for the plate-plate and sphere-plate configurations, and compare the average force to the analytical expressions in Fig.~\ref{FigSphereplatecomparison}. While the exact expressions are analytical, we evaluate them numerically and refer to their results as `numerical' (the finite-element results are `FEM'). The details of our numerical evaluation are described in the Supplementary Information. 

There is good agreement between the patch force for the finite-element and numerical results (Fig.~\ref{FigSphereplatecomparison}\textbf{A}), especially for the plate-plate simulation and at small separations. The distance-dependence of the force is stronger when the distance between surfaces is much larger than the patch size ($d \gg \ell$). The error bars indicate the standard deviation from independent patch realizations, and they are smaller than the data points until the minimum separation becomes large. At this point, in the limit $d\gg\ell$, the model becomes limited by the finite geometry size. A similar error occurs in the numerical evaluation of the analytical models if the coordinates of the evaluation points are not spread sufficiently far on the sphere and plate domains: The results from non-plane geometries tend towards those obtained from the plate-plate geometry. Nonetheless, our FEM model reasonably captures the force over a large range of separations.   

The force-distance scaling is shown in Fig.~\ref{FigSphereplatecomparison}\textbf{B}. It shows that the force converges to the model of patchless surfaces (i.e., as in a capacitor) in the limit $d\ll \ell$, with $F \propto 1/d$ ($n=1$) for the sphere-plate case\cite{Kim2008,deMan2009} and $F \propto 1/d^2$ ($n=2$) for the plate-plate case~\cite{Bressi2002,Zou2013}. In the limit $d\gg \ell$, the force-distance scalings converge to $F \propto 1/d^3$ ($n=3$) in the sphere-plate case and $F\propto 1/d^4$ ($n=4$) in the plate-plate case, which are the results expected from a distribution of electric dipoles on each surface~\cite{Speake2003}. Our FEM model replicates this reasonably well in the plate-plate case, but suffers from the finite size of the simulation domain in the sphere-plate case. 

In the limit $d\ll \ell$, the majority of the electrostatic energy exists between patches on opposite surfaces, and there is comparatively little interaction between patches on the same surface. That can be seen by schematically drawing the electric field lines from patches on a surface, Fig.~\ref{FigSphereplatecomparison}\textbf{C}, which closely resemble the field lines seen within a plate capacitor. Vice-versa, for $d \gg \ell$, the majority of the electrostatic energy exists between patches on the same surface. That is why care is needed when creating a finite-element mesh for patches if $d \gg \ell$, as small changes in the mesh can obscure the force associated from the change in $d$. The field around the border between patches on the same surface is similar to a dipole field, which is why the force-distance scaling reproduces that of a surface of dipoles~\cite{Speake2003} in this limit. 

\begin{figure}
\includegraphics[width = 0.5\textwidth]{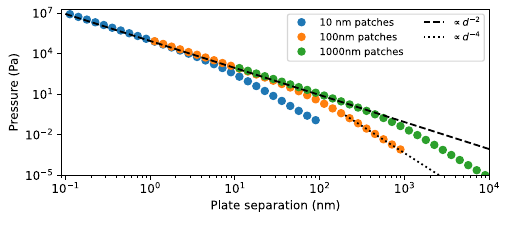}
\caption{Patch pressure between two flat plates for various patch sizes (colors) and plate separations, evaluated by our FEM method. In the limit of large patches, $d \ll \ell$, all curves follow the (patchless) capacitor case (dashed black line). In the limit of small patches, $d \gg \ell$, the pressure is proportional to $d^{-4}$.}
\label{Fig_various_scales}
\end{figure}

It may be of interest to estimate an upper bound to the patch force rather than explicitly modeling the entire experiment. In the next section, we show that the strongest possible patch force between two surfaces a distance $d$ apart occurs if those surfaces are flat plates. Between two plates, the upper bound of patch force is described by the patchless capacitor model, as shown in Fig.~\ref{Fig_various_scales}. This bound is the electrostatic pressure between two capacitor plates, $P = \frac{1}{2}\varepsilon_0 \frac{2\sigma_V^2}{d^2}$, where $\sigma_V$ is the standard deviation of the normal distribution of patches, \SI{100}{\milli\volt} in this case. This bound is reached in the large-patch limit, $d \ll \ell$, as long as the lateral dimensions of the plates are much larger than $\ell$ (i.e., there are many patches). 

\begin{figure*}[ht!]
\includegraphics[width = \textwidth]{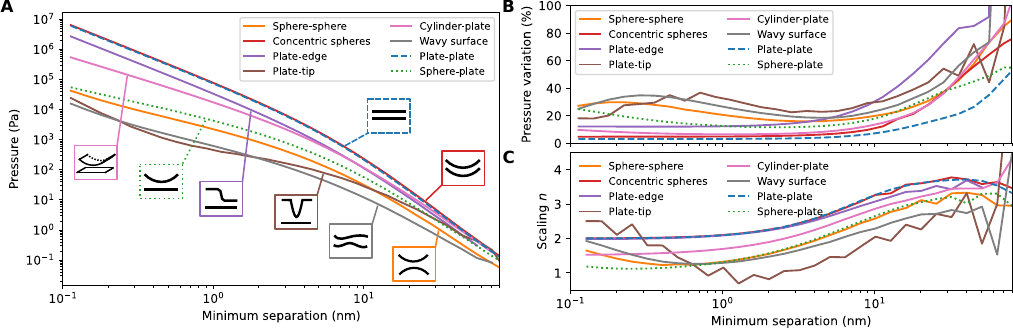}
\caption{The pressure exerted by potential patches between surfaces of various geometries. All curves are averages of $15$ random patch textures. All spherical surfaces follow a radius of \SI{250}{\nano\meter}, the edge has a step height $2\ell$, and the tip has a Lorentzian shape with height $\ell$ and a half-width at half-maximum of $\ell$. The wavy surface is the identical to the one shown in Fig.~\ref{FigSimulationschematic}.}
\label{FigVariousgeometries}
\end{figure*}

\subsection{Patch force in other geometries}
Having established that our finite-element model reproduces previous results when $d\ll \ell$ and $d\simeq \ell$, we now apply it to different geometries. We define two parametric surfaces with parameters $s_1$ and $s_2$ that span $[0,1]$, using $x = s_1$, $y = s_2$ and $z = f(s_1,s_2)$ (see the Supplementary for example expressions for $f(s_1,s_2)$). We find the minimum separation between the surfaces $z_\mathrm{bottom}$ and $z_\mathrm{top}$ by using the derivatives of the expression $f$ with respect to $x$ and $y$. This way we can compare the patch force at equal minimum separations regardless of the geometry. Furthermore, we restrict all simulations laterally to have a square domain with side length $L = $~\SI{340}{\nano\meter} and we compute patch pressures based on the area of this square.

All geometries we have explored yield a monotonic function of the minimum separation $d$, as shown in Fig.~\ref{FigVariousgeometries}\textbf{A}. However, we do not include geometries with surfaces indenting into one another such as those of Ref.~\cite{Tang2017}, where non-monotonic behavior has been observed. The largest patch pressure is found in the plate-plate and concentric-spheres geometries, as every point on one surface is at the same (minimal) distance from the other surface in the limit $d \ll \ell \ll R$ (radius $R = 250$~\si{\nano\meter}). Conversely, the plate-tip geometry has the smallest patch pressure, because there is only a small fraction of the total patch surface that is close to the plate. In between these extremes, there are the plate-edge, cylinder-plate, sphere-plate, and sphere-sphere geometries (from maximum to minimum pressure). In the limit of large $d$, the pressure in all cases tends to the $1/d^4$ dependence of the plate-plate geometry, which is a limitation of the finite size of the FEM geometry.

It is worthwhile to look how much the patch pressure varies between different patch textures in the same geometry, shown in Fig.~\ref{FigVariousgeometries}\textbf{B}. The lower limit is given by the number of patches we simulate, as it scales with $1/\sqrt{n}$. The plate-plate, concentric-spheres and plate-edge geometries show the least variation in pressure between the different realizations, while the plate-tip geometry shows the largest variation. In the latter, a large fraction of the patch interaction energy is present in a small volume at the apex of the tip, where it is not only sensitive to a single patch but also most sensitive to changes in the FEM mesh. 

In Fig.~\ref{FigVariousgeometries}\textbf{C} we show the pressure-distance scaling $n$ for the various geometries. With the exception of the plate-tip geometry, all geometries fall in between the plate-plate scaling ($n = 2$) and sphere-plate scaling ($n = 1$) in the large-patch limit. In the small-patch (or large $d$) limit, the FEM simulations converge to the plate-plate scaling due to the limited size of the FEM model geometry, and become noisy due mesh and other numerical effects. 

\subsection{Patch force with KPFM and AFM data}
\begin{figure}[ht!]
\includegraphics[width = 0.5\textwidth]{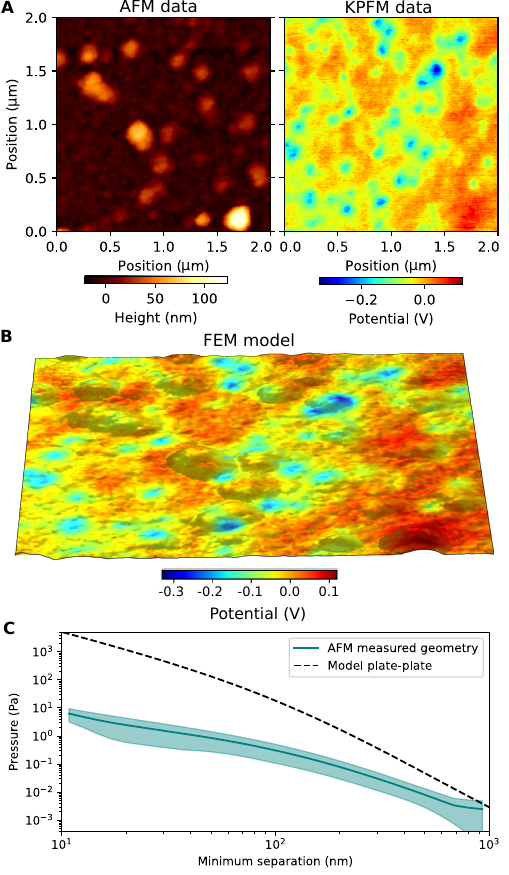}
\caption{\textbf{A}: Measurement of an Al surface, with height from a regular AFM measurement (left panel) and potential from a KPFM measurement (right panel). \textbf{B}: COMSOL model of the combined AFM and KPFM measurements, to be used as the bottom or top surface in our model for evaluation of the patch force. \textbf{C}: Pressure exerted by potential patches based on AFM and KPFM measurements. The shaded area indicates the spread between various $2\times 2$~\si{\micro\meter\squared} measurement areas close by on the same sample. The patch pressure in geometries with roughness is much less than the pressure between two perfectly plates (analytical model, dashed line).}
\label{FigAFMbasedgeometry}
\end{figure}
Our FEM model can import height data measured by an AFM to evaluate patch pressure in realistic situations. The parametric surface(s) shown in Fig.~\ref{FigSimulationschematic} can be defined as an interpolation between the data points of an AFM scan, to reproduce a surface with roughness. Furthermore, if the potential has been measured in the same surface area, by KPFM, it can also be included in the simulation by using this data instead of a randomly generated Voronoi pattern of potential, as shown in Fig.~\ref{FigAFMbasedgeometry}. 

We have scanned multiple $2\times 2$~\si{\micro\meter\squared} areas of an evaporated Aluminium film of thickness \SI{40}{\nano\meter} that has undergone multiple fabrication steps as detailed in Ref.~\cite{deJong2025}. The height and potential data were scanned simultaneously using the KPFM mode of a commercial AFM (Oxford Instruments Jupyter XR), shown in Fig.~\ref{FigAFMbasedgeometry}\textbf{A}. Both datasets were imported into COMSOL and used as input to generate the FEM model shown in Fig.~\ref{FigAFMbasedgeometry}\textbf{B}. The top and side surfaces of the model are not shown, for visual clarity. 

The pressure between surfaces where both height and potential are measured is shown as a function of distance in Fig.~\ref{FigAFMbasedgeometry}. It scales less strongly with distance than patches of the same potential energy applied to perfectly flat plates (dashed line), instead it more closely resembles the force-distance scaling of the tip-plate geometry. The height data shows multiple geometrical features comparable in height to the minimum separation distance, so it seems likely that the potential at these asperities dominates the interaction, which leads to the tip-plate behavior. Past discussions about the experimentally determined distance scaling of patch pressure already concluded that small-scale imperfections in experimental geometries (roughness, scratches) can cause force-distance scalings to deviate from those expected in ideal geometries \cite{Kim2008,Decca2009}. Our FEM simulations further support this point.

\subsection{Minimizing potential}
\begin{figure}
\includegraphics[width = 0.5\textwidth]{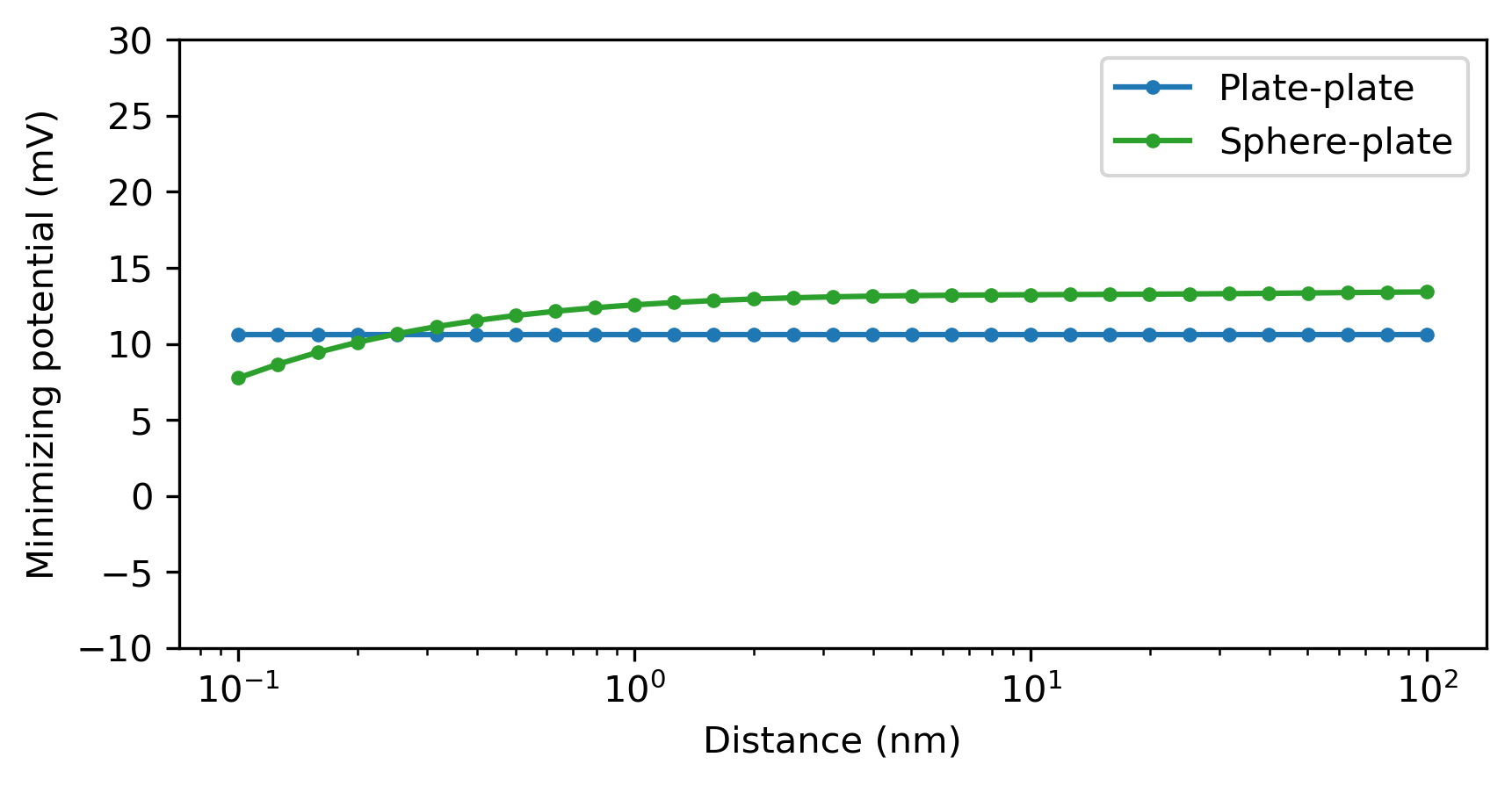}
\caption{Minimizing offset potential between two flat plates (blue) and a sphere-plate (green) as a function of distance. In the sphere-plate case, the minimizing potential is distance dependent because the single patch closest to the plate dominates the interaction at small separations. The patches on all surfaces are normally distributed with $\sigma_V = 100$~\si{\milli\volt}.}
\label{FigMinimizingpotential}
\end{figure}

Experiments sensitive to potential patches minimize their contribution by applying an offset potential which minimizes the electrostatic force (e.g., Ref.~\cite{Kim2008,Behunin2012}). We show in Fig.~\ref{FigMinimizingpotential} that our FEM model can extract the minimizing potential for any of the patch textures or geometries. We apply an offset potential, sum it with the patch texture, and solve the electrostatic problem as before. We find the minimizing potential by varying the offset potential and fitting a quadratic function, since the force $F \propto V^2$. In the case of perfectly flat plates (Fig.~\ref{FigMinimizingpotential}, blue) the minimizing potential is not dependent on distance $d$: All patches always contribute equally to the force regardless of distance. But if part of the geometry is curved, the closest patches contribute more for smaller $d$ so the minimizing potential varies as a function of distance. Since experimental geometry tends to be rough and have some curvature, the minimizing potential should vary as a function of distance if there are potential patches.

\section{Conclusion}
We have developed and provided (see \hyperlink{Paravailability}{Data availability}) an FEM-model to evaluate the parasitic contribution from electrostatic patches in various 3D geometries. Our calculation results match analytic and approximate results when they are available, in e.g., plate-plate and sphere-plate geometries, and we provide an straightforward explanation for the force-distance scaling in the limit of large and small patches. We have extended FEM-based calculations of the patch force in geometries for which analytic solutions are challenging or not available. We have furthermore shown that measured topography and potentials, from AFM and KPFM respectively, can be used as input for our FEM model to estimate the patch contribution in realistic experimental conditions. 
Generally, the patch contribution can be minimized by choosing materials with small patch sizes (patch size $\ell \ll d$, the distance between surfaces), suggesting thin-film materials with typically smaller grains may be preferred over bulk materials. 

Our model allows us to compute the parasitic contribution from potential patches in any experimental geometry, which will improve estimations of the parasitic force contribution to Casimir experiments~\cite{Speake2003,deJong2025} and other small-force sensing experiments. A possible future extension of our model should focus on simulating fluctuations of patch potentials, which are more relevant to some experiments than static force values. Other works have already investigated temporal fluctuations in some specific geometries~\cite{Robertson2006,Pollack2008,Dubessy2009,Ke2023,Vitale2024}, and we expect our FEM model will be helpful to estimate the force fluctuations in more varied experimental geometries. 

\vspace{0.25cm}
\paragraph*{Data availability}\mbox{}\\
\hypertarget{Paravailability}{All} data, simulations, measurement and analysis scripts in this work are available at \urlstyle{same}\url{https://doi.org/10.5281/zenodo.13255584}.

\vspace{0.25cm}
\paragraph*{Acknowledgments}\mbox{}\\
M.J. and L.M. acknowledge the Aalto Scientific Computing team for their support. L.M. acknowledges funding from the Strategic Research Council at the Academy of Finland (Grant No. 338565). This work was supported by the Finnish Quantum Flagship (FQF) funded by the Research Council of Finland (project number project number 358877, Aalto University).


%

\clearpage

\onecolumngrid
\setcounter{figure}{0}
\renewcommand{\thefigure}{S\arabic{figure}}
\setcounter{equation}{0}
\renewcommand{\theequation}{S\arabic{equation}}
\section{Supplementary Information}

\subsection{Expressions for parametric surfaces}
We list the expressions used for the parametric surfaces in Table \ref{TableParametricsurfaces}. We use some geometrical parameters: $L$ is the side length of the simulation domain, $R$ is the radius of curvature, and for simplicity we use $\tilde{s}_1 = s_1 - 0.5$, $\tilde{s}_2 = s_2 - 0.5$. 
\begin{table}
\begin{tabular}{l|l|l}
Geometry & $z_\mathrm{bottom}$ & $z_\mathrm{top}$ \\
\hline
Plate-plate & $0$ & $0$ \\
Sphere-plate & $0$ & $R\left(1 - \sqrt{1 - \left(\frac{L}{R}\sqrt{\tilde{s}_1^2 + \tilde{s}_2^2} \right)^2}\right)$ \\
Sphere-sphere & -$R\left(1 - \sqrt{1 - \left(\frac{L}{R}\sqrt{\tilde{s}_1^2 + \tilde{s}_2^2} \right)^2}\right)$ & $R\left(1 - \sqrt{1 - \left(\frac{L}{R}\sqrt{\tilde{s}_1^2 + \tilde{s}_2^2} \right)^2}\right)$ \\
Concentric spheres & $R\left(1 - \sqrt{1 - \left(\frac{L}{R}\sqrt{\tilde{s}_1^2 + \tilde{s}_2^2} \right)^2}\right)$ & $R\left(1 - \sqrt{1 - \left(\frac{L}{R}\sqrt{\tilde{s}_1^2 + \tilde{s}_2^2} \right)^2}\right)$ \\
Cylinder-plate & $0$ & $R\left(1 - \sqrt{1 - \left(\frac{L}{R}\tilde{s}_1 \right)^2}\right)$ \\
Edge-plate & $0$ & $0.1~\mathrm{erf}(20\tilde{s}_1)$ \\
Tip-plate & $0$ & $-0.001 (\tilde{s}_1^2 + \tilde{s}_2^2 + 0.1^2)^{-1}$ \\
Wavy surface & $0.05 \sin (2 \pi s_1) + 0.03 \cos (2 \pi s_2)$ & $0.05 \cos (2 \pi s_1) + 0.04 \sin (2 \pi s_2)$
\label{TableParametricsurfaces}

\end{tabular}
\end{table}

\subsection{Numerical evaluation of the analytical model for plate-plate patches}\label{Secnumericalmodel}
To calculate the electrostatic interaction from patches on two plates, we follow the method of Ref.~\cite{Speake2003}. The expression for the electrostatic energy per unit surface (\si{\joule\per\meter\squared}) in the system of plate separated by distance $d$ is
\begin{equation}
\mathcal{E}_\mathrm{pp} = - \frac{\varepsilon_0}{8\pi^2} \iint  k\frac{(C_{11} + C_{22}) \cosh (k d)- 2 C_{12}}{\sinh (k d)} d^2 \vec{k} .
\end{equation}
We use 2D reciprocal coordinates $\vec{k} = [k_x, k_y]^T$, and $k^2 = k_x^2 + k_y^2$. The quantities $C_{11}(\vec{k})$, $C_{22}(\vec{k})$, and $C_{12}(\vec{k})$ are the spatial spectral densities of the potential distributions $V_1$ and $V_2$ on plates 1 and 2 respectively. They are defined by $C_{ij} = \left\langle V_i(\vec{k}) V_j(-\vec{k}) \right\rangle$. The vacuum permittivity is denoted by $\varepsilon_0$. The electrostatic pressure $P$ can then be computed as \cite{Speake2003}
\begin{equation}
P = -\frac{\varepsilon_0}{8\pi^2} \iint \frac{k^2}{\sinh^2(kd)} \big( C_{11} + C_{22} - 2 C_{12} \cosh (kd)\big) d^2 \vec{k},
\label{Eqplatepatchforce}
\end{equation}
as long as the plate separation is much smaller than the lateral dimensions (negative values denote an attractive pressure).  

Different statistical descriptions of patches have been proposed, e.g., the sharp-cutoff model that assumes that all patches fall between a minimum and maximum size~\cite{Speake2003}, and the quasi-local model that sums over all (circular) patches that contain any two points on the plate~\cite{Behunin2012}. The statistical properties of patches in a given material can be extracted from KPFM measurements~\cite{Garrett2015}. We can generate different patch textures, realizations of random patches, and insert them into Eq.~\eqref{Eqplatepatchforce} through the spatial correlations, $C_{11}$ and $C_{22}$ for patches on the same plate and $C_{12}$ between the patches on different plates. We follow the model of Ref.~\cite{Ke2023} by generating the patches using Voronoi diagrams of randomly distributed points, which is a well-known way of modeling polycrystalline solids. 
Notwithstanding the discussions about the precise statistical properties of $C_{11}$, $C_{22}$, and $C_{12}$~\cite{Speake2003,Behunin2012,Ke2023}, it is generally accepted that patches are uncorrelated at large distances, and that patches on different surfaces are uncorrelated, i.e.,
\begin{equation}
\begin{aligned}
\lim_{\lvert r \rvert\rightarrow \infty} C_{ii}(\vec{r}) &= 0 \\
\langle C_{12} \rangle = \langle C_{21}\rangle &= 0.
\label{Eq_patch_statistical_properties}
\end{aligned}
\end{equation}

\begin{figure}
\includegraphics[width = 0.5\textwidth]{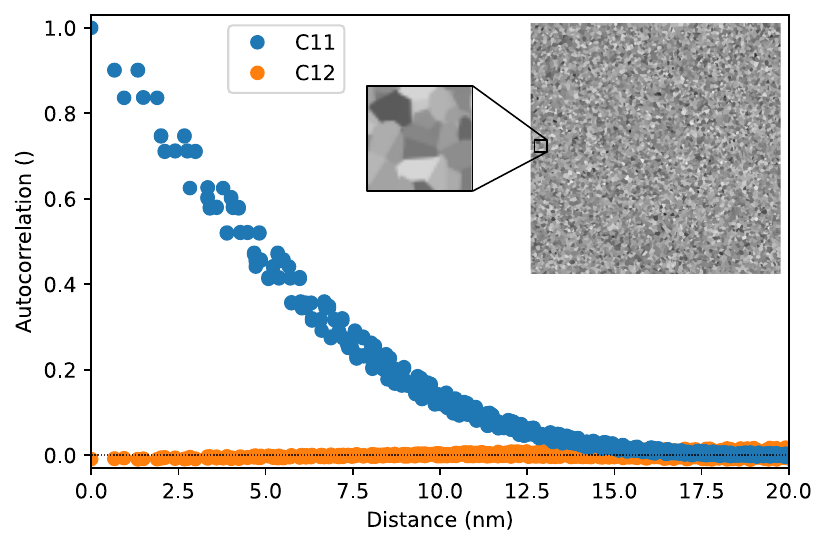}
\caption{Autocorrelation function of the Voronoi diagram used to model patches, with the patches and their (greyscale) potential shown as an inset. This model has $10,000$ patches in a $1\times 1$ \si{\micro\meter\squared} area, for an average patch size of \SI{10}{\nano\meter}. The zoom-in covers a $50 \times 50$ \si{\nano\meter\squared} area and shows the discretization (pixelation) used for the numerical analysis.}
\label{Figstatisticalpatchmodel}
\end{figure}

We use scipy~\cite{Virtanen2020} to generate the Voronoi diagram for a number of randomly distributed points. We then assign a random potential value to each region (grain) by filling the area with a random greyscale color using matplotlib. We save the resulting image and subsequently use the pixelation as a spatial discretization of the domains. We subtract the average greyscale value of all the pixels to obtain a Gaussian distribution centered around zero. The correlations $C_{11}$, $C_{22}$ and $C_{12}$ can be directly computed from the image matrices using scipy. An example Voronoi diagram is shown in Fig.~\ref{Figstatisticalpatchmodel} (inset), with the zoom-in showing the pixelation. The autocorrelation of potential (blue dots) shows a decreasing trend qualitatively similar to the quasi-local patch model of~\cite{Behunin2012}. At large distances, the correlation between patches on the same plate tends to zero, and the cross-correlation between patches on the same plate, shown with orange dots, is close to 0 everywhere. Thus this method of generating patches is in agreement with Eq.~\eqref{Eq_patch_statistical_properties}.

We evaluate the force exerted between two plates with patches by generating a patch texture with $100,000$ patches with a characteristic size $\ell = 10$~\si{\nano\meter} and potentials following a normal distribution of standard deviation \SI{100}{\milli\volt}. 


\subsection{Numerical evaluation of the analytical model for sphere-plate patches}
\begin{figure}
\includegraphics[width = \textwidth]{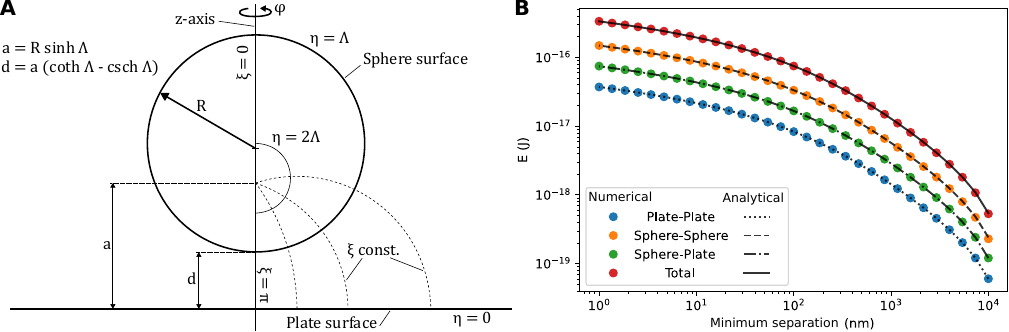}
\caption{\textbf{A}: Schematic of the bispherical coordinate system used in the sphere-plate configuration, adapted from \cite{Behunin2012a}. \textbf{B}: Electrostatic energy in the sphere-plate system without patches, for $V_\mathrm{p} = 1$~\si{\volt} and $V_\mathrm{s} = -2$~\si{\volt}. Our numerical evaluation method follows the analytical expression closely.}
\label{Sphereplateschematic}
\end{figure}

We base our numerical evaluation of patches in the sphere-plate geometry on the work of Ref.~\cite{Behunin2012a}. Their analytical expression gives the solution for the potential for the boundary value problem where the patches are taken as fixed potentials on the sphere and plate domains. They use the bispherical coordinate system, which is shown schematically in Fig.~\ref{Sphereplateschematic}\textbf{A}. The patch potentials are defined on the domain $\Omega$ such that $V(\Omega) \equiv V(\xi,\phi)$.  The radius of the sphere $R$ and the minimal distance between sphere and plane $d$ relate to the quantity $\Lambda$ as \cite{Behunin2012a}
\begin{equation}
\cosh \Lambda = 1+d/R \qquad \text{and} \qquad R\sinh \Lambda = \sqrt{(d+R)^2-R^2} = a.
\end{equation}
which does not depend on the integration measure, which is the solid angle $\int d\Omega = \int_0^\pi d\xi \int_0^{2\pi} d\phi \sin \xi$ in the bispherical coordinate system. 

Ref.~\cite{Behunin2012a} gives the expressions for the interaction energy in the sphere-plate geometry, split into sphere-sphere ($E_\mathrm{sp}^{s,s}$), sphere-plate ($E_\mathrm{sp}^{s,p}$), and plate-plate ($E_\mathrm{sp}^{p,p}$) terms,
\begin{equation}
\begin{aligned}
E_\mathrm{sp}^\mathrm{s,s} &= \frac{\varepsilon_0}{2\pi} \sum_{n=1}^\infty \int d\Omega \int d \Omega^{'} \frac{e^{-\Lambda n} R \sinh \Lambda V_\mathrm{s}(\Omega) V_\mathrm{s}(\Omega^{'} ) \mathcal{G}(e^{-2\Lambda n},\cos \gamma)}{(\cosh \Lambda - \cos \xi)(\cosh \Lambda - \cos \xi')}, \\
E_\mathrm{sp}^\mathrm{s,p} &= -\frac{\varepsilon_0}{\pi} \sum_{n = 0}^\infty \int d\Omega \int d\Omega^{'} \frac{e^{-(n+1/2) \Lambda} R \sinh \Lambda V_\mathrm{s}(\Omega) V_\mathrm{p}(\Omega^{'}) \mathcal{G}(e^{-2 (n + 1/2) \Lambda}, \cos \gamma)}{(\cosh \Lambda - \cos \xi)(1 - \cos \xi')}, \\
E_\mathrm{sp}^\mathrm{p,p} &= \frac{\varepsilon_0}{2\pi} \sum_{n=1}^\infty \int d\Omega \int d\Omega^{'} \frac{e^{-\Lambda n} R \sinh \Lambda V_\mathrm{p}(\Omega) V_\mathrm{p}(\Omega^{'} ) \mathcal{G}(e^{-2\Lambda n},\cos \gamma)}{(1-\cos \xi) (1 - \cos \xi^{'})}.
\end{aligned}
\label{EQ_sphereplate_totalenergy}
\end{equation}
In the above expression, the subscripts $s$ and $p$ denote the integration over the sphere and plate domain, respectively. The generation function $\mathcal{G}(t,u)$ is used to express an infinite sum over the orders of the Legendre polynomial $P_l^k$,
\begin{equation}
\sum_{l=0}^\infty \lambda_l^2 t^l P_l(u) = \frac{1-10t^2+t^4+4t(1+t^2)u}{4(1+t^2-2tu)^{5/2}} \equiv \mathcal{G}(t,u),
\end{equation}
where the addition theorem for Legendre polynomials is used such that
\begin{equation}
\begin{aligned}
P_l (\cos \gamma) &= \sum_{k = -l}^l (-1)^k e^{ik(\phi-\phi')} P_l^k (\cos \xi)P_l^{-k}(\cos \xi'), \\
\cos \gamma &= \cos \xi \cos \xi' + \sin \xi \sin \xi' \cos (\phi - \phi').
\end{aligned}
\end{equation}
Finally we use the shorthand $\lambda_l = l + 1/2$. 

Ref.~\cite{Behunin2012a} also gives the interaction energy of the sphere-plate geometry and the electrostatic force in the absence of patches are described by the expressions 
\begin{equation}
\begin{aligned}
E_\mathrm{sp}^\mathrm{eq} &= 2\pi\varepsilon_0 (V_\mathrm{s}-V_\mathrm{p})^2 R \sinh \Lambda \sum_{n = 1}^\infty \frac{1}{\sinh \Lambda n}, \\
F_\mathrm{sp}^\mathrm{eq} &= - 2\pi \varepsilon_0 (V_\mathrm{s} - V_\mathrm{p})^2 \sum_{n = 1}^\infty \frac{\coth \Lambda - n\coth \Lambda n}{\sinh \Lambda n},
\label{Eq_sphereplate_energy_nopatches_total}
\end{aligned}
\end{equation}
where $V_\mathrm{s}$ and $V_\mathrm{p}$ are the potentials of the sphere and plate respectively. As noted in Ref.~\cite{Behunin2012a}, the divergent plate self-energy has been subtracted from the above expression but the finite, distance-independent sphere self energy is still present.  

We now describe our method to numerically compute the electrostatic energy in the sphere-plate geometry. We first pick a set of random points on both surfaces that act as patch centers, and assign a normally-distributed potential value to each of the patches. We then generate a set of evaluation points distributed across the sphere and plate, and compute $\cos \gamma$ for each of those points (this does not depend on the sphere-plate separation in bispherical coordinates). We are generally interested in the change of energy with separation $d$, so we loop over this variable. In the loop, we shift the coordinates of the patch centers on the sphere, so that they are at the correct distance from the plate. Then we calculate the local value $V(\xi,\phi)$ for both the sphere and plate potentials, by finding which patch center each evaluation point is closest to. In Cartesian coordinates, this is easier than in bispherical coordinates, so we make use of the coordinate transform
\begin{equation}
x = \frac{a \sin \xi \cos \phi}{\cosh \eta - \cos \xi} \qquad y = \frac{a \sin \xi \sin \phi}{\cosh \eta - \cos \xi} \qquad z = \frac{a \sinh \eta}{\cosh \eta - \cos \xi},
\end{equation}
where $\eta = 0$ on the plate and $\eta = \Lambda$ on the sphere. After this we compute the sums over $n$, where we truncate $n \leq 300 d^{0.4}$, which is an empirical bound where higher order terms contribute negligibly to the sum. Finally, we multiply the total sum over $n$ with the potential values and integrate over all evaluation points. To save computational power, we consider different realizations of the random potentials assigned to each patch. Since the geometry is the same, we calculate its effect only once, and average over many random patch potential realizations.

We validate our numerical method by assigning all patches to the same potential value, $V_\mathrm{p} = 1$\si{\volt} and $V_\mathrm{s} = -2$\si{\volt}, and computing the energy for different sphere-plate distances $d$. We plot the energies obtained from the different terms of our numerical computation of Eg.~\eqref{EQ_sphereplate_totalenergy} and the total analytical expression (Eq.~\eqref{Eq_sphereplate_energy_nopatches_total}) in Fig.~\ref{Sphereplateschematic}\textbf{B}. There is excellent agreement between our numerical method (colored circles) and the analytical expressions (black lines, expressions for the individual terms can be found in Ref.~\cite{Behunin2012a}).


\end{document}